\documentclass[onecolumn, 10pt]{article}
\usepackage[a4paper, margin=1.0in]{geometry}
\usepackage{amsmath,amssymb}
\usepackage{graphicx}
\usepackage{float}
\usepackage{diagbox}
\usepackage[format=hang,labelfont=bf,textfont=small,singlelinecheck=false,justification=raggedright,margin={12pt,12pt},figurename=Fig.]{caption}
\usepackage{titlesec}
\usepackage{mathtools}
\usepackage{textcomp}
\usepackage{float}
\usepackage{hyperref}
\usepackage{subcaption} 
\usepackage[export]{adjustbox}
\usepackage{etoolbox}
\usepackage{authblk}

\makeatletter
\def\affiliation#1{\gdef\@affiliation{#1}}
\def\abstract#1{\gdef\@abstract{#1}}
\def\graphabst#1{\gdef\@graphabst{#1}}
\def\keywords#1{\gdef\@keywords{#1}}
\def\corresp#1{\gdef\@corresp{#1}}

\newcommand{\MakeTitle}{
  \newpage
  \null
  \vskip 2em%
  \begin{center}%
  \Large \@title\par
  \vskip 1em%
  \large \@author
  \end{center}
  \noindent\@affiliation\par
  \vskip 1em%
  \noindent\@corresp\par
  \vskip 1em%
  \noindent\@abstract\par
  \vskip 1em%
  \noindent\@keywords\par
}
\makeatother

\makeatletter
\patchcmd{\@maketitle}{\raggedright}{\centering}{}{}
\patchcmd{\@maketitle}{\raggedright}{\centering}{}{}
\makeatother

\newcommand*{\TitleFont}{%
      \usefont{\encodingdefault}{\rmdefault}{}{n}%
      \fontsize{18}{12}%
      \selectfont}

\titleformat{\section}
  {\normalfont\fontsize{10}{11}\bfseries}{\thesection.}{2pt}{}
  \titlespacing*{\section}{0pt}{12pt}{6pt}

\titleformat{\subsection}
  {\normalfont\fontsize{10}{10}\bfseries}{\thesubsection.}{2pt}{}
  \titlespacing*{\subsection}{0pt}{6pt}{0pt}
  
\titleformat{\subsubsection}
  {\normalfont\fontsize{10}{10}\bfseries}{\thesubsubsection.}{2pt}{}
  \titlespacing*{\subsubsection}{0pt}{6pt}{0pt}
  
\normalsize
\title{\TitleFont Unexplored aspects of a variational principle in electrostatics}

\author[1]{Kolahal Bhattacharya}
\author[2]{Debapriyo Syam}
\affil[1]{Homi Bhabha Centre for Science Education (TIFR), Mumbai, India}
\affil[2]{Guest Faculty, Centre for Astroparticle Physics and Space Science, Bose Institute, Kolkata, India}
\date{}
\corresp{$^{\ast}$kolahal.bhattacharya@hbcse.tifr.res.in}

\abstract{\textbf{Abstract}: 
The electrostatic field magnitude can play a role in a variational principle similar to the role of the index of refraction in geometrical optics, allowing the determination of the electric field lines. This was identified in the context of the grounded conducting sphere image problem. Assuming the knowledge of the magnitude of the electric field, in this paper we validate this principle for the general case and explore some of its consequences. 
}

\begin{document}
\maketitle

\section{Introduction}
Variational principle constitutes one of the most powerful methods of physical theories. The most popular of them is, of course, Hamilton's principle of stationary action~\cite{goldstein2011classical}. In geometrical optics, one would use Fermat's principle~\cite{landau2013classical} that asserts that light rays follow the path about which the optical path length is stationary. Many major results of geometrical optics can be directly derived from this principle.

One of the methods of solving boundary value problems in electrostatics is the method of images~\cite{jackson1999classical}. Its name is perhaps due to the apparent similarity between the image formation by a plane mirror and the placement of an image charge in the case of an infinite grounded conducting plane. Though not widely recognised, there is a deeper similarity even in other cases, for example the grounded conducting sphere image problem~\cite{bhattacharya2011analogy}. In this case, one finds analogues of the mirror image and object distance relations as well as magnification formulae. This observation led to the discovery of a variational principle $\delta\int_{A}^{B}E\ ds=0$ in electrostatics~\cite{bhattacharya2013novel} like Fermat's principle in geometrical optics, where $A$ and $B$ are two fixed points lying on the same electrostatic field line. Here it is assumed that the magnitude of the electric field is known as a function of position. Due to the conservative nature of the field, the integral $\int_A^B{\bf E}\cdot{d {\bf s}}$ between two fixed points $A$ and $B$ is stationary, irrespective of the curve. However, the proposed variational principle leads to the curve between $A$ and $B$ superimposed with the local direction of the field.

Though this principle was originally contemplated to explain the analogy with virtual image formation in mirror-optics, 
it can have important implications. For example, recently it has been applied to demystify the nonlocality problem in the Aharonov-Bohm effect~\cite{bhattacharya2021demystifying} that has resolved a fundamental paradox of quantum physics. Even in the classical domain, there are other yet unexplored illuminating aspects of this principle that can shed light on the properties of electrostatic fields.
We shall discuss some of these in this paper. We shall start by giving a proof of the variational principle in two dimensions in the next section. Further, we shall construct an electrostatic Lagrangian from this principle in the following section and observe the results when the system is invariant under spatial translation or rotation. Finally, we shall conclude with an application of this principle to derive Green's differential equation that relates the rate of change of the electrostatic field, in the direction perpendicular to an equipotential surface, to the principal radii of curvature of the surface.
\section{Proof of $\delta\int_{A}^{B}{E}\ ds=0$ about a field line (in two dimensions)}
For pedagogical reasons, in this section we shall first obtain in two dimensional space the Euler-Lagrange equation that is compatible with the variational principle. Afterwards, the validity of this equation will be demonstrated using an independent approach. 
Let us assume that the electrostatic field ${\bf E}$ is confined to two dimensions (say, in the $x-y$ plane). Then, the variational principle between the fixed points can be expressed as:
\begin{equation}\label{eq1}
    \delta\int_{A}^{B}{E}\ ds=\delta\int_{A}^{B} E(x,y(x)) \sqrt {1+y'^2}dx=0,
\end{equation}
where $E$ denotes the magnitude of the electrostatic field, and prime (') denotes the derivative with respect to $x$. Eq.\eqref{eq1} can be treated as a stationary action principle with the factor $E(x,y(x))\sqrt{1+y'^2}$ identified as the corresponding Lagrangian function. The Euler-Lagrange equation for this Lagrangian is expressed as:
\begin{equation}\label{eq2}
    \frac{d}{dx}\left[E\frac{y'}{\sqrt{1+y'^2}}\right]-\frac{\partial E}{\partial y}\sqrt{1+y'^2}=0.
\end{equation}
Therefore, the task in hand is to prove that the left hand side of Eq.\eqref{eq2} indeed equals zero in a manner independent of the proposed variational principle. 
Now, electric field vector ${\bf E}$ is related to the potential $\phi$ by ${\bf E}=-\nabla\phi$. Let us represent the field line through the point $(x, y)$ by $ \psi(x,y)=k$ which implies that $\nabla\psi$ is perpendicular to the field lines directed along $\nabla\phi$. On the $x-y$ plane, this condition can be expressed as: 
\begin{equation}\label{eq3}
    \frac{\partial\psi}{\partial x}\frac{\partial\phi}{\partial x}+\frac{\partial\psi}{\partial y}\frac{\partial\phi} {\partial y}=0.
\end{equation}
Along the field line $d\psi=0$, so we have:
\begin{equation}\label{eq4}
    0 = \frac{\partial\psi}{\partial x} dx + \frac{\partial\psi} {\partial y} dy.
\end{equation}
Using Eq.\eqref{eq3} and Eq.\eqref{eq4}, we find that:
\begin{equation}\label{eq5}
    \frac{dy}{dx}=y'=\frac{\left(\frac{\partial\phi}{\partial y}\right)}{\left(\frac{\partial\phi}{\partial x}\right)}.
\end{equation}
This result can also be obtained by drawing an electric field line on the $x-y$ plane and noting that the gradient at any point $dy/dx=E_y/E_x$. Now, writing $E=\left|\nabla\phi\right|$, the operand of ${d}/{dx}$ in Eq.\eqref{eq2} can be simplified as:
\begin{align}\label{eq6}
  |\nabla\phi|\frac{y'}{\sqrt{1+y'^2}}&=\sqrt{(\frac{\partial \phi}{\partial x})^2+(\frac{\partial\phi}{\partial y})^2}\frac{\frac{\left(\frac{\partial\phi}{\partial y}\right)}{\left(\frac{\partial\phi}{\partial x}\right)}}{\sqrt{(\frac{\partial \phi}{\partial x})^2+(\frac{\partial\phi}{\partial y})^2}}\frac{\partial\phi}{\partial x}\nonumber\\
  &=\frac{\partial\phi}{\partial y}.
\end{align}
Thus, the first term on the left hand side of Eq.\eqref{eq2} equals:
\begin{equation}\label{eq7}
    \frac{d}{dx}\left(\frac{\partial\phi}{\partial y}\right)=\frac{\partial^2\phi}{\partial x\partial y}+\frac{\partial^2\phi}{\partial y^2}\frac{dy}{dx}.
\end{equation}
On the other hand, the second term on the left hand side of Eq.\eqref{eq2} equals:
\begin{align}\label{eq8}
    \frac{\partial|\nabla\phi|}{\partial y}\sqrt{1+y'^2}
    =\frac{\frac{\partial\phi}{\partial x}\frac{\partial^2\phi}{\partial y\partial x}+ \frac{\partial\phi}{\partial y}\frac{\partial^2\phi}{\partial y^2}}{\sqrt{(\frac{\partial \phi}{\partial x})^2+(\frac{\partial\phi}{\partial y})^2}}\cdot\frac{\sqrt{(\frac{\partial \phi}{\partial x})^2+(\frac{\partial\phi}{\partial y})^2}}{\frac{\partial\phi}{\partial x}}
    =\frac{\partial^2\phi}{\partial x\partial y}+\frac{\partial^2\phi}{\partial y^2}\frac{dy}{dx},
\end{align}
assuming continuity of the second partial derivatives for the electrostatic potential and taking into account the expression for ${dy}/{dx}$ from Eq.\eqref{eq5}. From Eq.\eqref{eq6}, Eq.\eqref{eq7} and Eq.\eqref{eq8}, we find that the left hand side of Eq.\eqref{eq2} indeed equals zero. Hence, the original proposition is true when the points $A$ and $B$ lie on an electrostatic field line. A coordinate-free proof, on the other hand, is given in~\cite{bhattacharya2013novel}.

\section{An analogue of Newton's second law}\label{Newton2}
The strong similarity of the proposed principle with the Fermat's principle $\delta\int_A^B n\ ds=0$, where $n$ denotes the index of refraction, calls for a careful comparison of the properties of the light rays and those of the electric field lines. In three dimensions, the light ray path is specified by a differential equation which is nothing but the Euler-Lagrange equation from the Fermat's principle. Let us define a stepping parameter $a$, such that the position of a point on a ray can be expressed by: ${\bf r}=[x(a),y(a),z(a)]$. As $a$ changes, ${\bf r}$ moves smoothly over the path of the ray. Additionally, if we denote ${\bf r'}={d{\bf r}}/{da}$, then the differential equation of the light ray path can be expressed as~\cite{evans1986f}:
\begin{equation}\label{eqPreNewton}
    \nabla n\left|{\bf r'}\right|=\frac{d}{da}\left(\frac{n \bf r'}{|\bf r'|}\right).
\end{equation}
If the stepping parameter $a$ is chosen as the arc length $s$, we have $\left|{\bf r'}\right|=\left|{d{\bf r}}/{ds}\right|=1$ and Eq.\eqref{eqPreNewton} reduces to
\begin{equation}\label{eq10}
    \nabla n=\frac{d}{ds}\left(n\frac{d{\bf r}}{ds}\right).
\end{equation}
Evidently, this is the standard form of the equation which can be derived from the eikonal equation in geometrical optics~\cite{born2013principles}. In this context, we notice that a similar Euler-Lagrange equation can be derived for the electrostatic fields, from $\delta\int_{A}^{B} E\ ds=0$ along a field line~\cite{bhattacharya2013novel},
\begin{equation}\label{eq9}
    \nabla E=\frac{d}{ds}\left(E\frac{d{\bf r}}{ds}\right).
\end{equation}
Thus, as far as the differential equations of light rays and of the field lines of the static fields are concerned, the scalar fields $n$ and $E$ behave in a similar manner. Now, both Eq.\eqref{eq10} and Eq.\eqref{eq9} are coupled nonlinear equations. If, however, we choose the parameter $a$ in Eq.\eqref{eqPreNewton}, such that $\left|{d{\bf r}}/{da}\right|=n$, then Eq.\eqref{eqPreNewton} reduces to a much simpler form, that resembles Newton's second law of motion~\cite{evans1986f}:
\begin{equation}\label{eq11a}
 \frac{d^2{\bf r}}{da^2}=\nabla\left(\frac{n^2}{2}\right).
\end{equation}
In a similar manner, if in electrostatics the stepping parameter $a$ is chosen in such a way that 
\begin{equation}\label{eqES-refractive-index}
    \frac{d{\bf r}}{da}={\bf E},
\end{equation}
implying that $\left|{d{\bf r}}/{da}\right|=E$, then the spatial evolution of the electric field lines is governed by
\begin{equation}\label{eq12a}
 \frac{d^2{\bf r}}{da^2}=\nabla\left(\frac{E^2}{2}\right).
\end{equation}
Following Evans et al.~\cite{evans1986f}, one can identify the analogue of potential energy as $-{E^2}/{2}$, the analogue of mass as 1, and the analogue of total energy as $\mathcal{E}=(1/2)|{d{\bf r}}/{da}|^2-{E^2}/{2}=0$. From Eq.\eqref{eq12a}, we can readily calculate the differential equation of the electric field lines if the function $E=|\nabla\phi|$ is known and vice versa. For example, if we know that the electric field near a charged plate is constant, then Eq.\eqref{eq12a} asserts that the equation of the electric field line is ${d^2{\bf r}}/{da^2}= 0$, i.e., ${\bf r}$ is linear in $a$. Hence, eliminating $a$ from $x(a), y(a)$ etc. we easily find that the field lines $y(x),z(x)$ etc. are nothing but straight lines. It is clear that this equation can also be used to calculate the equations of the fringing field lines if the electric field magnitude is known. For example, one can deduce the shapes of the field lines at the edges of a parallel plate capacitor, from the information of potential or field. A general expression of the electric field near a conductor, dependent on local surface parameters was derived in~\cite{bhattacharya2016dependence}.

\section{Consequences of invariance of electrostatic systems under spatial translation and rotation}
Blaker et al~\cite{blaker1974application} applied Noether's theorem to geometrical optics to deduce the invariant quantities corresponding to continuous transformations that leave the system in the original configuration. In fact, this can be accomplished in a straightforward manner without an explicit reference to the Noether's theorem just by using the Euler-Lagrange equations, when the symmetry information can be incorporated into Lagrangian function. We shall demonstrate this for the electrostatic systems using the following systems with either (a) translation symmetry, or (b) rotational symmetry. Let us first note that the variational principle can be expressed as:
\begin{equation}\label{eqELag}
    \delta\int E\left(\frac{ds}{da}\right)\ da=0,
\end{equation}
-where $a$ denotes the stepping parameter introduced in the last section. Eq.\eqref{eqELag} corresponds to a Lagrangian $E\left( {ds}/{da}\right)$. In terms of generalised orthogonal curvilinear coordinates $(q_1, q_2, q_3)$ and scale factors $(h_1, h_2, h_3)$ -where $h_i={\partial s}/{\partial q_i}$, one can express the line element $ds$ as
\begin{equation}
    ds=\sqrt{h_1^2(dq_1)^2+h_2^2(dq_2)^2+h_3^2(dq_3)^2},
\end{equation}
and hence, the electrostatic Lagrangian becomes
\begin{equation}\label{GenLagrangian}
    E(q_1, q_2, q_3) \sqrt{h_1^2q_1'^2+h_2^2q_2'^2+h_3^2q_3'^2},
\end{equation}
where $q_i'={dq_i}/{da}$ for $i=(1,2,3)$. In this notation, the Euler-Lagrange equation becomes
\begin{equation}\label{EuLag}
    \frac{d}{da}\left(\frac{\partial L}{\partial q_i'}\right)-\frac{\partial L}{\partial q_i}=0.
\end{equation}
From Eq.\eqref{EuLag}, it is clear that if the Lagrangian $L$ is independent of $q_i$, then the generalised momentum $p_i= {\partial L}/{\partial q_i'}$ is an invariant, independent of $a$. Also we note that in the Cartesian coordinate system: $(q_1,q_2,q_3)= (x,y, z)$, $h_1=h_2 =h_3=1$ and Eq.\eqref{GenLagrangian} reduces to:
\begin{equation}\label{Eq19}
    E(x,y,z)\sqrt{\left(\frac{dx}{da}\right)^2+\left(\frac{dy}{da}\right)^2+\left(\frac{dz}{da}\right)^2}=E^2,
\end{equation}
since ${dx}/{da}=E_x$, etc. This is consistent with our earlier discussion in section~\ref{Newton2}. There we defined the analogue of the kinetic energy as $T=(1/2)\left|{d{\bf r}}/{da} \right|^2={E^2}/{2}$ and the analogue of the potential energy as $-{E^2}/{2}$. Thus, the Lagrangian is: $T-V=E^2$.

\subsection{Translation symmetry}
\begin{figure}[H]
\centering
\begin{subfigure}{0.5\textwidth}
    \includegraphics[width=7.5cm, height=6.0cm]{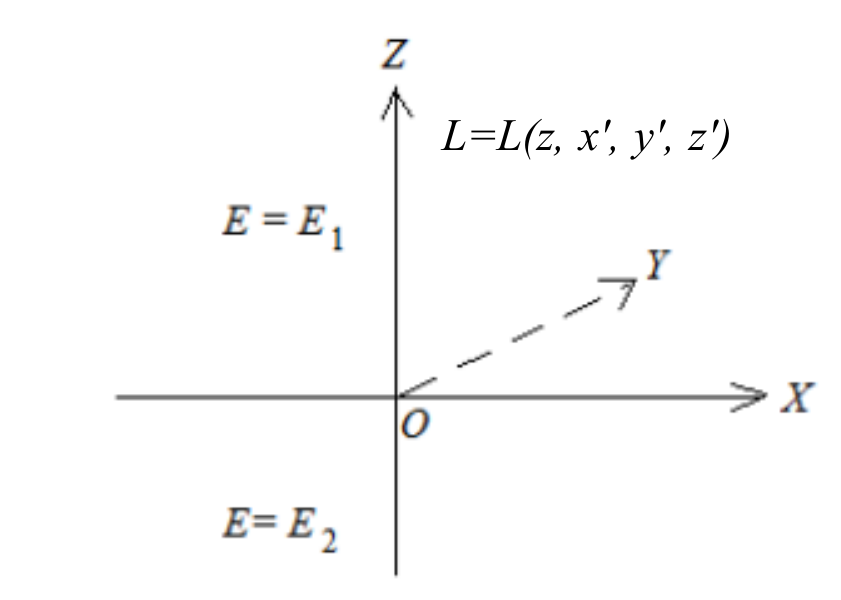}
    \caption{}
    \label{fig1a}
\end{subfigure}%
\hspace{-1.0cm}
\begin{subfigure}{0.5\textwidth}
\centering
    \includegraphics[width=7.5cm, height=6.0cm]{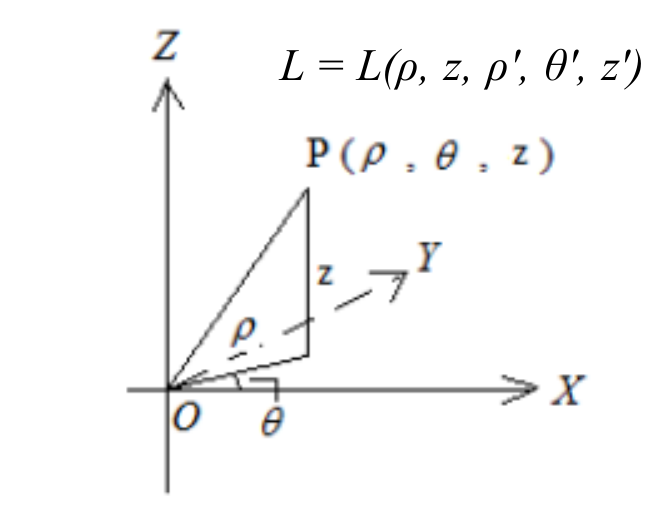}
    \caption{}
    \label{fig1b}
\end{subfigure}
\caption{
Diagrams corresponding to (a) a translationally symmetric electrostatic Lagrangian $L(z,x',y',z')$ which is independent of $x$ and $y$, and (b) a rotationally symmetric electrostatic Lagrangian $L(\rho,z,\rho',\theta',z' )$ which does not depend on $\theta$, the azimuthal angle.}
\label{}
\end{figure}

Let us consider two distinct values and directions of the electric field with magnitudes, $E_1$ and $E_2$, in the Cartesian system of coordinates: $(q_1,q_2,q_3)=(x,y,z)$, as shown in Fig.~\ref{fig1a}. Then, $h_1=h_2 =h_3=1$. We choose $E=E_1$ for $z>0$ and $E=E_2$ for $z<0$. Here we are considering a Lagrangian of the form $L=E(z(s)) \sqrt{x'(s)^2+y'(s)^2+z'(s)^2} $, and the object is to solve the equations of motion for the $x^i(s)$. Since L does not depend on $x$ or $y$, therefore we have
\begin{equation}
\frac{\partial L}{\partial x'}=E\frac{x'}{\sqrt{x'^2+y'^2+z'^2}}=E\frac{dx}{ds}=E_x=\rm{constant},
\end{equation}
using Eq.\eqref{Eq19}. If the electric field vector lies in the $x-z$ plane, then the above result means that the tangential component of the field vector has a constant value across the surface. 
Clearly, this derivation, based on the Lagrangian formulation of electrostatics, is consistent with the usual approach that begins with $\oint{\bf E}\cdot d{\bf r}=0$ and in which the line integral is evaluated along a closed path that skims infinitesimally above and below the boundary between two media. 

\subsection{Rotation symmetry}
Suppose the electric field has an axial symmetry. Let the $z$ axis be this symmetry axis. Then, the use of cylindrical polar coordinate system $(\rho,\theta,z)$ is more convenient (as shown in Fig.~\ref{fig1b}). The symmetry about z axis translates to:
$E=E(\rho,z)$ which is independent of $\theta$. The factor 
\begin{equation}
    \frac {ds}{da}=\sqrt{\left(\frac{d\rho}{da}\right)^2+\rho^2\left(\frac{d\theta}{da}\right)^2+\left(\frac{dz}{da}\right)^2},
\end{equation}
which is also independent of $\theta$. Thus, the Lagrangian $L$ is independent of $\theta$ as a whole. Hence, from Eq.\eqref{EuLag}, we have:
\begin{equation}\label{Eq22}
    \frac{\partial L}{\partial\theta'}=E(\rho,z)\frac{\rho^2\theta'} {\sqrt{\rho'^2+\rho^2\theta'^2+z'^2}}=\rm{constant}.
\end{equation}
But $\sqrt{\rho'^2+\rho^2\theta'^2+z'^2}=\sqrt{x'^2+y'^2+z'^2 }\equiv{E}$ and $\rho^2\theta'=(xy'- x'y)$. Thus, Eq.\eqref{Eq22} reduces to:
\begin{equation}\label{SHI}
    xy'-x'y=xE_y-yE_x=\rm{constant}.
\end{equation}
The left hand side of the above equation is the counterpart of the Smith-Helmholtz function in optics. The preceding exercise shows that this function is an invariant in axially symmetric system in electrostatics as well. Examples of such systems include dipoles, a thin charged wire, a charge placed in front of a grounded conducting sphere, etc. Differentiating Eq.\eqref{SHI} w.r.t. $a$, we can obtain a general relation between the $x$ and $y$ coordinates of the electrostatic field lines for axially symmetric systems:
\begin{equation}
    x\frac{d^2y}{da^2}=y\frac{d^2x}{da^2},
\end{equation}
valid for any given $z$ plane. 

\section{Application to prove Green's differential equation}
At every point on a smooth two dimensional surface, there are two principal radii of curvature $R_1$ (larger) and $R_2$ (smaller). The principal directions corresponding to these radii are mutually orthogonal. So, at that point, the surface normal planes in the principal directions are perpendicular to each other and both are orthogonal to the local tangent plane. Now, Green's differential equation (also known as Thomson's theorem) relates the normal derivative of electrostatic field $E$ in a charge free region near an equipotential surface to the mean curvature $\kappa=(1/2)\left({1}/{R_1}+ {1}/{R_2}\right)$ of the surface at the same point
\begin{align}\label{eqGreenThomson}
\frac{dE}{dn}=\left(\frac{dE}{ds}\right)_{along\ \hat{n}}= -\left(\frac{1} {R_1}+\frac{1}{R_2}\right)E,
\end{align}
where $dn$ denotes a differential length element in the direction of the outward normal at a point on the equipotential surface. This equation has been proved by many authors~\cite{bakhoum2008proof, estevez1985power, pappas1986differential}, often by lengthy approaches and differential geometric techniques. Here we present a simple proof of the theorem using Eq.\eqref{eq9}. Let us note that:
\begin{align}\label{eqGreen}
    \nabla\cdot{\bf E}  &=\nabla\cdot(E{\hat{\bf n}})\nonumber \\
                        &=\nabla E\cdot{\hat{\bf n}} + E\nabla\cdot{\hat{\bf n}}\nonumber\\
                        &=\frac{d{\bf E}}{dn}\cdot{\hat{\bf n}} + E(2\kappa),
\end{align}
-where we have used Eq.\eqref{eq9} in evaluating the first term, and the definition $\kappa=\nabla\cdot\hat{\bf n}/2$ (see, for example, Eq.(4.3.28) in pp. 178 of~\cite{pozrikidis2016fluid}). Just outside the equipotential, where Laplace's equation holds, we have $\nabla\cdot{\bf E}=0$. Therefore, from Eq.\eqref{eqGreen}, we have:
\begin{align}
    \frac{dE}{dn}&=\frac{d\bf E}{dn}\cdot{\hat{\bf n}}\nonumber\\
                 &=-2\kappa E\nonumber\\
                 &=-\left(\frac{1}{R_1}+\frac{1}{R_2}\right)E.
\end{align}
Thus, the fact that the mean curvature at a point on a surface is given by $\kappa=\nabla\cdot\hat{\bf n}/2$ leads to perhaps the simplest proof of this equation. We can also say that if at a point $\nabla\cdot{\bf E}={\rho}/{\epsilon_0}$, where $\rho$ is the source charge density and $\epsilon_0$ denotes to the permittivity of vacuum, the general relation will be:
\begin{equation}\label{eqGreenGeneral}
    \frac{dE}{dn} + \left(\frac{1}{R_1}+\frac{1}{R_2}\right)E=\frac{\rho}{\epsilon_0}.
\end{equation}
An example of the above relation can be found in the case of a uniformly charged sphere of radius $R$ with charge density $\rho={3Q}/{4\pi R^3}$. The electrostatic field at a point $r(<R)$ is given by $E={\rho r}/{3\epsilon_0}$. The normal direction is along the radial direction $\hat{r}$ which shows ${dE}/{dn}\equiv{dE}/{dr}$. Again, the equipotential surface at this point is a sphere of radius $r$ and has a mean curvature of $\kappa=1/r$. Note that, here, $R_1=R_2=r$. Evaluating the sum on the left hand side of Eq.~\eqref{eqGreenGeneral}, we find that the right hand side equals ${\rho}/{\epsilon_0}$. 

\section{Summary and discussion}
In this paper, we explored some aspects of the variational principle in electrostatics observed in~\cite{bhattacharya2013novel} in the context of the method of images. Significant new insights were gained when we discovered that the spatial evolution of the electrostatic field lines could be expressed as an analogue of Newton's second law of motion. We also deduced a general relation between the transverse coordinates of the electric fields for an axially symmetric electrostatic system. Finally, we showed that the equation of the electric field lines can be used to give a simple proof of Green's differential equation in electrostatics. It is anticipated that the use of this principle will lead to more interesting physical ideas in the future.



\end{document}